\documentclass[aip,apl,reprint]{revtex4-2}

\usepackage{graphicx}
\usepackage{xcolor}
\usepackage{amsmath}
\usepackage{dcolumn}
\usepackage{bm}

\begin{document}

\preprint{}

\title{\textbf{Stabilization of Ferroelectric Hafnia and Zirconia through Y$_2$O$_3$ doping}}%

\author{Li Yin}
\email{lyin@carnegiescience.edu}
\author{Cong Liu}
\author{R.~E.~Cohen}
\affiliation{Extreme Materials Initiative, Earth and Planets Laboratory, Carnegie Institution for Science \\ 5241 Broad Branch Road NW, Washington, District of Columbia 20015, USA}

\date{\today}

\begin{abstract}
We investigate the possible stabilization of ferroelectricity in bulk Y$_2$O$_3$-doped hafnia and zirconia. We use density functional theory (DFT) with large random supercells of hafnia and zirconia and study the relative phase stability of the centrosymmetric cubic and monoclinic phases compared with the polar orthorhombic phase. We find that Y$_2$O$_3$-doping stabilizes the polar ferroelectric phase over the monoclinic baddeleyite phase in both hafnia and zirconia.
\end{abstract}

\maketitle
Solid solution of Y$_2$O$_3$ in zirconia (ZrO$_2$) or hafnia (HfO$_2$) increases the stability of the cubic fluorite structure by about 2000 K \cite{scottPhaseRelationshipsZirconiayttria1975, duThermodynamicAssessmentZrO2YO151991, wuThermodynamicAssessmentHfO2YO151997}. Beautiful crystals of cubic zirconia are a commodity and Y-stabilized zirconia (YSZ) also has many industrial uses. Pure zirconia (ZrO$_2$) is centrosymmetric, and exists as monoclinic baddeleyite (space group $P2_1/c$) under ambient conditions, transforms to tetragonal $P4_2/nmc$ at approximately 1350 K, and to cubic fluorite $Fm\overline{3}m$ at about 2650 K. It has been extensively studied using DFT \cite{riccaComprehensiveDFTInvestigation2015} and machine learning ab initio molecular dynamics. Although the ground state of zirconia and hafnia is the centrosymmetric baddeleyite structure, hafnia and zirconia thin films show ferroelectricity \cite{bosckeFerroelectricityHafniumOxide2011, mullerFerroelectricitySimpleBinary2012}. Ferroelectricity has been observed experimentally in bulk Y-doped hafnia, and the polar phase was attributed to metastable formation due to fast quenching \cite{xuKineticallyStabilizedFerroelectricity2021}. Ferroelectric crystals up to 50 mm long were grown \cite{xuKineticallyStabilizedFerroelectricity2021, mikolajickFerroelectricityBulkHafnia2021}. Computations performed in these two papers did not find the ferroelectric phase to be stabilized, but this may be due to the small supercells of 48 atoms used there. Here we use large supercells and explore the relative stability of the cubic, tetragonal, and polar orthorhombic structures as functions of Y$_2$O$_3$ content. To maintain charge-balance there is one oxygen vacancy for every two substituted Hf or Zr. Throughout we give the percentage Y$_2$O$_3$ expressed as molar percentage Y replacement of Hf or Zr.

\begin{figure}[t]
    \centering
    \includegraphics[width=0.65\linewidth]{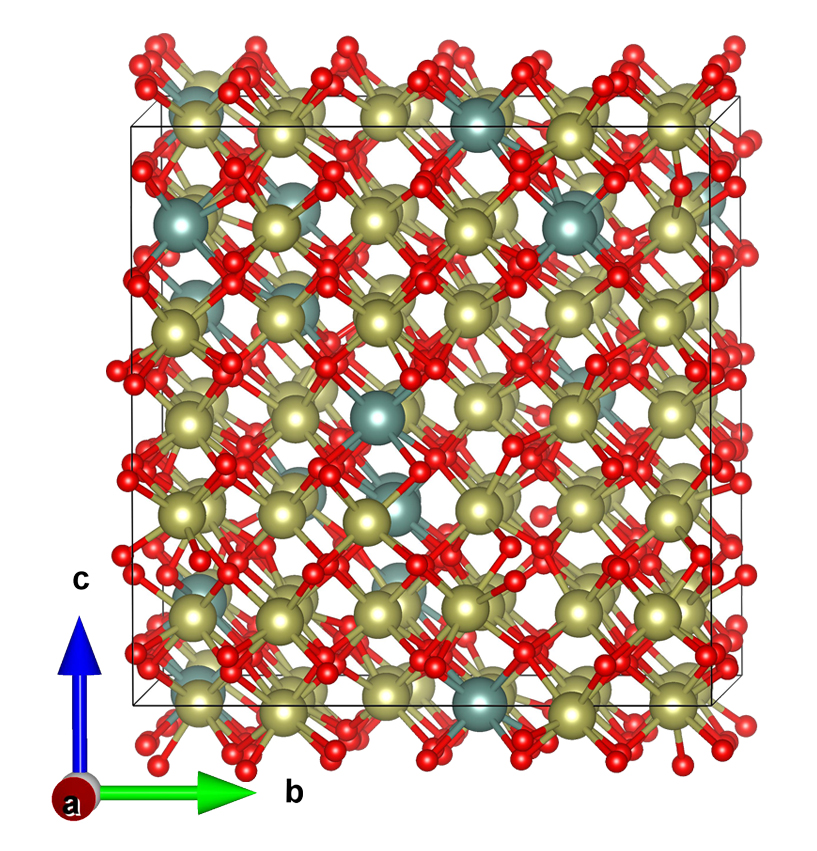}
    \caption{Structure of fully relaxed 14.8\% Y-doped hafnia in the orthorhombic polar phase. The blue, brown, and red balls represent Y, Hf, and O, respectively. The positions of the randomly distributed defects are generated using the SQS method, and then the structure is relaxed by minimizing forces and stresses. There is one oxygen vacancy for each two Y-dopants replacing Hf.}
    \label{fig:structure}
\end{figure}

We built a series of $3\times3\times3$ supercells of the 12-atom orthorhombic primitive cell (a total of 108 formula units) at different Y concentrations using the special quasi-random structures (SQS) method (Fig. \ref{fig:structure}) \cite{zungerSpecialQuasirandomStructures1990, vandewalleEfficientStochasticGeneration2013}. Our justification for using this approach is that at high temperatures, such as in the liquid phase, the atomic arrangement should be close to random, and with cooling the crystals do not order, since it involves diffusion of the highly charged Zr$^{4+}$ or Hf$^{4+}$ ions. Some oxygen redistribution is possible giving short-range ordering. 

\begin{figure*}[t]
    \centering
    \includegraphics[width=0.7\textwidth]{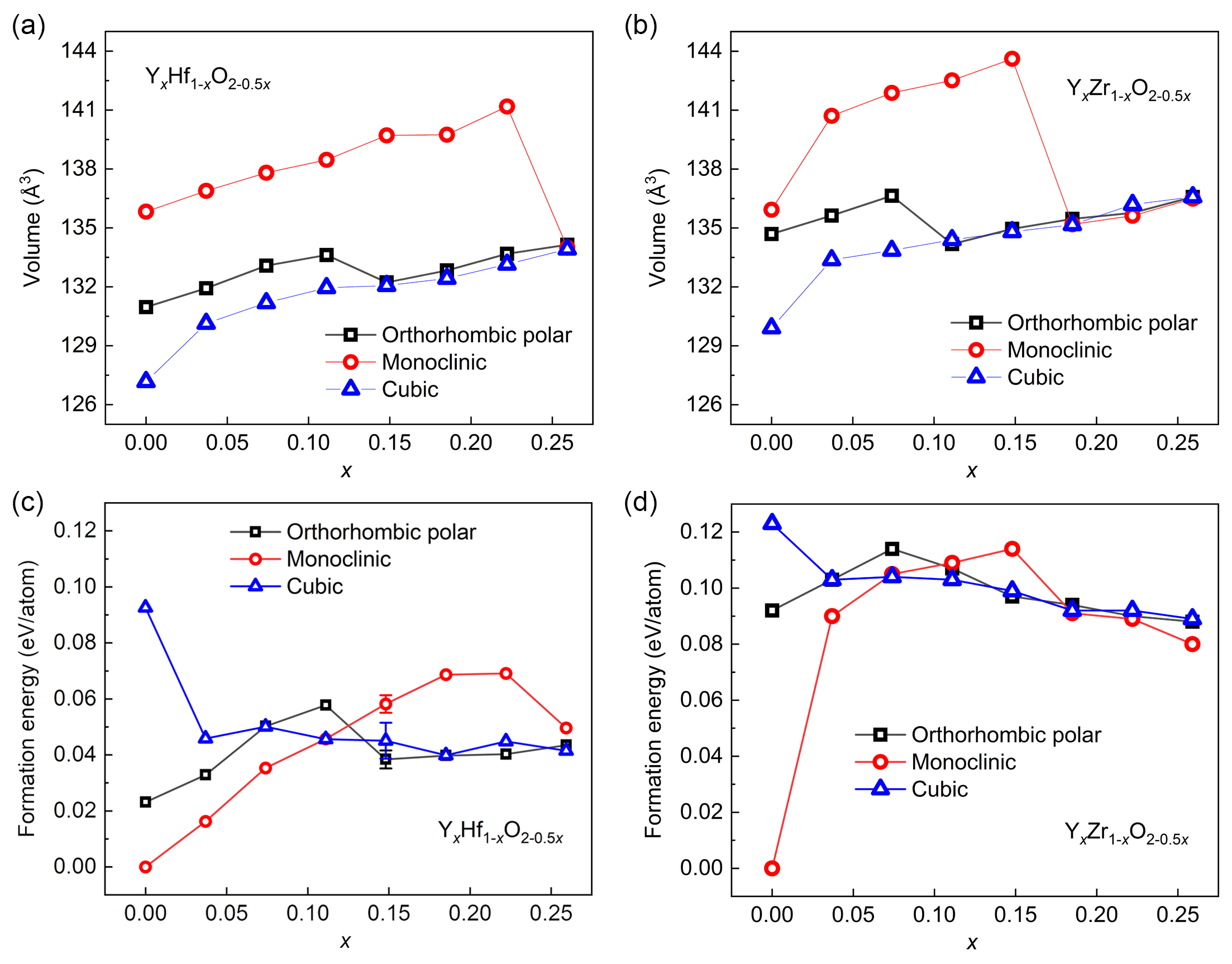}
    \caption{The volumes of Y$_2$O$_3$ doped HfO$_2$ (a) and ZrO$_2$ (b) and formation energies computed for Y$_2$O$_3$ doped HfO$_2$ (c) and ZrO$_2$ (d) in $3\times3\times3$ supercells. The error bar in (c) comes from the SQS cutoff tests.}
    \label{fig:figure1}
\end{figure*}

A different approach was used in Ref. \cite{xuKineticallyStabilizedFerroelectricity2021}, where all possible configurations in a very small supercell were exhaustively simulated, using the lowest energy configurations. In that work the lowest formation energy of 12.5\% Y-doped hafnia is 0.031 eV/atom in the ferroelectric phase, which is 0.026 eV/atom and 0.007 eV/atom lower than the formation energies in our 11.1\% and 14.8\% supercells, where we assume random distributions of dopants and vacancies. Expressed as k$_B$ T, where k$_B$ is the Boltzmann constant, the corresponding temperatures are 302 K and 81 K, so at high temperatures random arrangements as we assume seem highly probable. 

We also generated smaller $2\times2\times2$ supercells using the SQS method, and fully relaxed them by DFT. The energy differences and lattice constants computed with the $2\times2\times2$ supercells are consistent with the results in the $3\times3\times3$ supercells (Figs. S1, S2) for Y content up to 12.5\%. We find differences at higher concentrations (Figs. \ref{fig:figure1}(c) and S1), which shows that smaller supercells give inaccurate results at higher doping concentrations.

We computed the formation energies of doped hafnia and zirconia for the orthorhombic, monoclinic, and cubic phases with different Y$_2$O$_3$ concentrations (Fig. \ref{fig:figure1}). In each case we started with the fully symmetric pure HfO$_2$ or ZrO$_2$ structures, placed the defects obtained by SQS as described above, and then relaxed the structures until forces and stresses were zero. The defects break the symmetry in any finite supercell, but we refer to each by the symmetry of the pure starting structures. The formation energies are defined as

\begin{multline}
    E_{\text{form}}=(E(\text{Y}_x\text{Hf}_{1-x}\text{O}_{2-0.5x})-(1-x)E(\text{HfO}_2) \\
    -0.5xE(\text{Y}_2\text{O}_3))/3,
\end{multline}
where the monoclinic \textit{P}2$_1$/\textit{c} ground state of pure HfO$_2$ and cubic \textit{Ia}$\overline{3}$ ground state of pure Y$_2$O$_3$ are used \cite{xuElectronicStructuralOptical1997}. For both hafnia and zirconia, we find that the polar orthorhombic phase becomes more stable than the monoclinic and cubic phases at a Y molar concentration of 14.8\%. The formation energies of Y-doped zirconia at ferroelectric phase are higher than those in Y-doped hafnia, suggesting that the ferroelectric phase would be harder to form in zirconia compared with hafnia. 

We find a large change in volume in the polar phase in hafnia at a Y concentration of 14.8\% (Fig. \ref{fig:figure1}(a)), caused by the dramatically decreased lattice constant \textit{a} (see also Fig. S2). The fluorite lattice acquires a tetragonal strain with a small c/a of \(1.01\sim1.015\) for Y concentrations less than 11.1\%, but it becomes more cubic with Y content higher than 11.1\% (Fig. S2). This can be an artifact of our finite supercells--the infinite (or large) cubic solid solution would have cubic symmetry. The fluorite phase shows lower formation energy at 11.1\% Y-doped hafnia and zirconia at zero temperature. The polar orthorhombic phase exhibits lower formation energy from 14.8\% to 22.2\% Y-doped hafnia and in a small region around 14.8\% Y concentration in zirconia. The small formation energy difference between cubic and polar orthorhombic phases are discussed at the end of this paper.

For high Y concentration (\textit{x}=14.8\%, 18.5\%, or 22.2\%), the energy difference between polar and cubic phases is smaller than energy difference between polar and monoclinic (Figs. \ref{fig:figure1}(c)), so that baddeleyite is destabilized with high Y-contents, and the ferroelectric phase competes with the cubic phase at those high Y concentrations. The stabilization of cubic fluorite with Y$_2$O$_3$ is consistent with the known HfO$_2$-Y$_2$O$_3$ and ZrO$_2$-Y$_2$O$_3$ phase diagrams \cite{stacyYttriaHafniaSystem1975, stapperInitioStudyStructural1999, jacobsonThermodynamicModelingYO15ZrO22004, chenThermodynamicModelingZrO2YO152004, predithInitioPredictionOrdered2008}.

We also computed the phonon dispersion in Y-doped ferroelectric hafnia and ordered approximations of cubic zirconia (see Supplement). For hafnia phonon computations we used $2\times2\times2$ supercells for 6.25\% and 12.5\% Y-doped models.  For these compositions, as discussed above, agreement is good compared with the larger supercells. Long-range interactions were included using the computed effective charges (see Supplement) and the non-analytical correction \cite{giannozziInitioCalculationPhonon1991, gonzeInteratomicForceConstants1994}. 
\begin{figure}[b]
    \centering
    \includegraphics[width=0.95\linewidth]{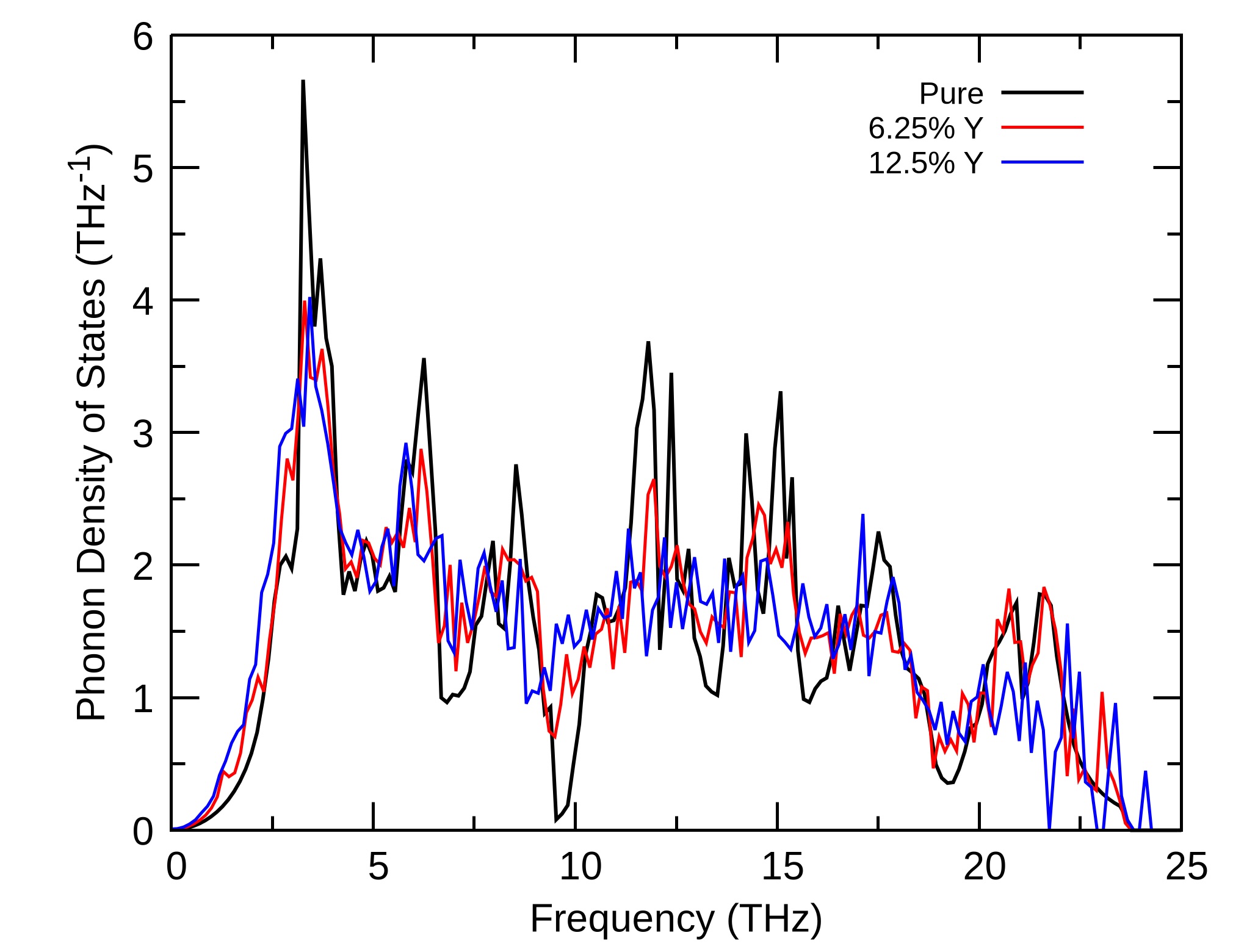}
    \caption{Computed phonon density of states in pure, 6.25\%, and 12.5\% Y-doped hafnia in $2\times2\times2$ supercell.}
    \label{fig:phonon}
\end{figure}
The Y-doped structures show no imaginary frequencies (Fig. \ref{fig:phonon}). We thus find that Y-doping in hafnia stabilizes the orthorhombic polar phase energetically and vibrationally as the ground state. No additional strains, defects, or appeal to kinetics is necessary.  

We computed piezoelectric constants in pure hafnia, 6.25\%, and 12.5\% Y-doped hafnia within the $2\times2\times2$ supercells. The computed elastic moduli, direct and converse piezoelectric tensors in pure hafnia (see Table \ref{tab:table1} and Supplement Table S1) are consistent with previous values \cite{duttaPiezoelectricityHafnia2021}. We find that piezoelectricity in Y-doped hafnia is modest at 0 K, but it will increase at temperatures approaching phase transitions as observed in other materials \cite{acostaPiezoelectricityRotostrictionPolar2016, acostaBaTiO3basedPiezoelectricsFundamentals2017}.

In hafnia thin films, the sign of the longitudinal piezoelectric coefficient is tunable via epitaxial strain \cite{chengTunableParabolicPiezoelectricity2024}. We find the sign of \textit{e}$_{33}$ or \textit{d}$_{33}$ is negative for Y concentration up to 12.5\%. The sign of \textit{e}$_{15}$ or \textit{d}$_{15}$  switches from negative to positive at Y dosage of 12.5\%.

\begin{table}[t]
\caption{\label{tab:table1}The total direct piezoelectric tensor \textbf{\textit{e}} (C/m$^2$) and converse piezoelectric tensor \textbf{\textit{d}} (pm/V) in  the ferroelectric phase of 0\% (pure), 6.25\%, and 12.5\% Y-doped hafnia at 0 K. The doping structure is in $2\times2\times2$ supercell. Indices are given in Voigt notation.
}
\begin{ruledtabular}
\begin{tabular}{lcccc}
\textrm{Index}&
\textrm{0\% Y in Ref \cite{duttaPiezoelectricityHafnia2021}}&
\textrm{0\% Y}&
\textrm{6.25\% Y}&
\textrm{12.5\% Y}\\
\colrule
e31	&-1.31	&-1.39	&-0.98	&-0.62\\
e32	&-1.33	&-1.34	&-1.17	&-0.69\\
e33	&-1.44	&-1.46	&-1.65	&-0.92\\
e15	&-0.20	&-0.23	&-0.003	&0.14\\
e24	&0.64	&0.70	&0.77	&0.76\\
\\				
d31	&-1.71	&-2.01	&-0.69	&-0.92\\
d32	&-1.77	&-1.65	&-1.46	&-0.86\\
d33	&-2.51	&-2.46	&-3.79	&-1.91\\
d15	&-2.03	&-2.44	&-0.03	&2.03\\
d24	&6.74	&7.42	&9.31	&10.2\\
\end{tabular}
\end{ruledtabular}
\end{table}

We computed the electric polarization of orthorhombic polar (\textit{Pca}2$_1$), monoclinic, and cubic hafnia at different Y doping concentrations using the Berry's phase approach based on modern theory of polarization \cite{king-smithTheoryPolarizationCrystalline1993, restaMacroscopicPolarizationCrystalline1994, restaDipoleCrystallitePolarization2021} and using Born effective charges. The electric polarization is defined as follows
\begin{equation}
\textbf{\textit{P}}=\textbf{\textit{P}}_{0} \pm n\textbf{\textit{P}}_q \pm \textbf{\textit{P}}_Z \\
    =\frac{|e|}{V}\sum_{i}Z_i^* \textbf{\textit{r}}_i \pm \frac{|e|}{V_{prim}}\textbf{\textit{R}}_{prim} 
    \pm \frac{|e|}{V}Z_i^*,
\end{equation}
\begin{equation}
\textbf{\textit{R}}_{prim}=n_{1}\textbf{\textit{a}}_{1}+n_{2}\textbf{\textit{a}}_{2}+n_{3}\textbf{\textit{a}}_{3},
\end{equation}
where $\textbf{\textit{P}}_{0}$ is initial polarization, $\textbf{\textit{P}}_{q}$ is the polarization lattice or quantum, $\textbf{\textit{P}}_Z$ is the Born effective charge polarization quantum (see beloow), \textit{e} is the electron charge, \textit{V} is the volume of unit cell, $Z^*$ is the Born effective charge, $\textit{\textbf{r}}_i$ denotes the Cartesian position of atom i, sum runs over all the atoms in unit cell, \textbf{\textit{R}} is lattice vector in primitive cell, $\textbf{\textit{a}}_{1}, \textbf{\textit{a}}_{2}, \textbf{\textit{a}}_{3}$ are the lattice basis vectors in primitive cell. Since Eq. (2) involves a sum over the entire system, one must carefully group canceling or nearly canceling contributions. The Born polarization quantum is introduced by the Born effective charge of atoms in the border, corner, or face sites of the cell, to obtain a meaningful result from an infinite sum involving both diverging positive and negative terms. 

We computed the effective charges in ferroelectric phase using density functional perturbation theory (DFPT) (see Supplement Table S2). Since the $3\times3$ effective charge tensors depend on site symmetry, phase, and atomic environment, we use average values here in order to obtain approximate but robust results for different compositions, supercells, and dopant arrangements. The average Born effective charges for Hf and O are 5.26 and -2.63, respectively. Using these Born effective charges to calculate the electric polarization in polar phase per formula (2) gives 0.58 C/m$^2$, close to the polarization we calculated via the direct Berry's phase computation value of 0.55 C/m$^2$. Our polarization agrees with that previously computed \cite{climaIdentificationFerroelectricSwitching2014, fanOriginIntrinsicFerroelectricity2019}. 

\begin{figure}[t]
    \centering
    \includegraphics[width=0.95\linewidth]{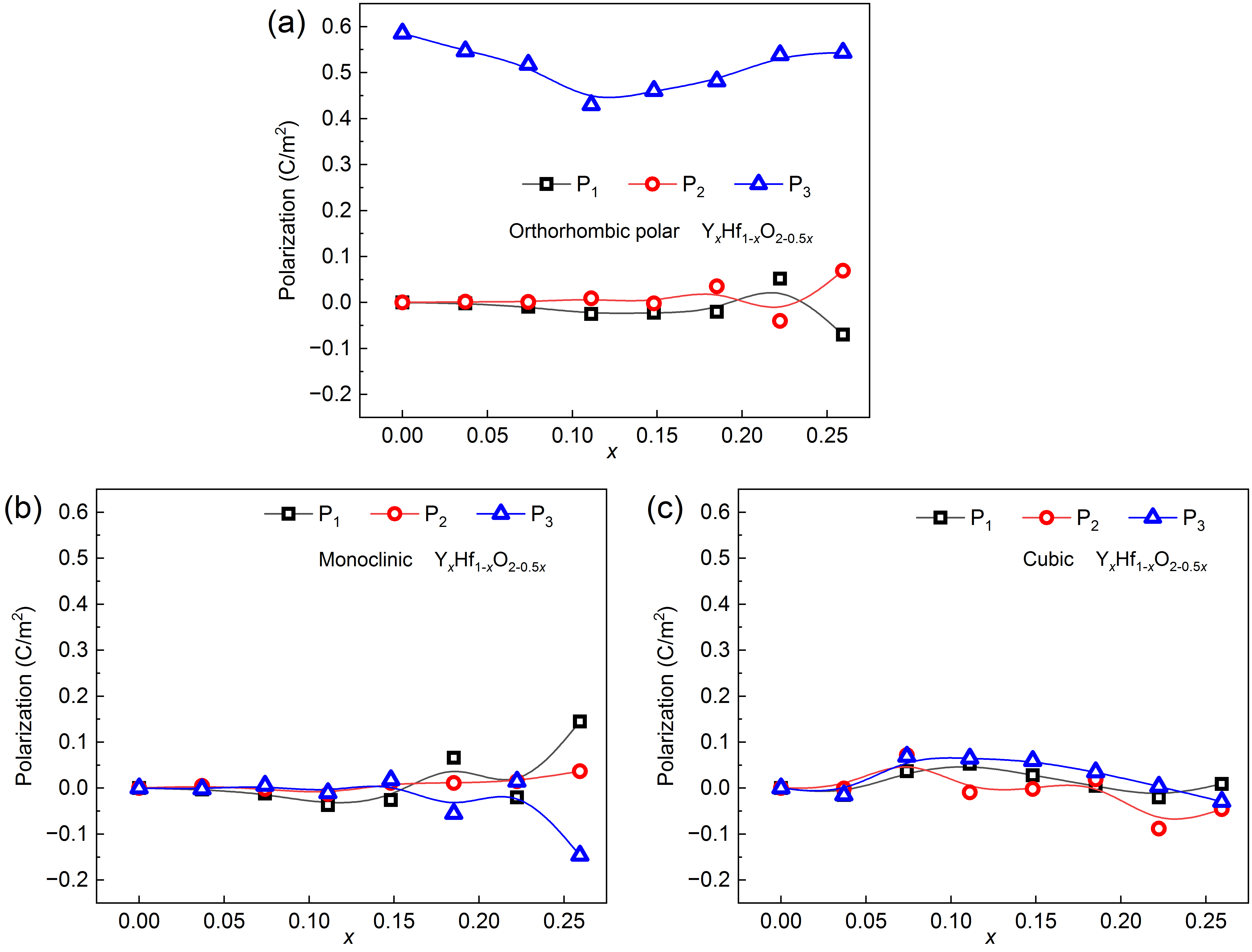}
    \caption{Decomposed electric polarization in a series of fully relaxed Y-doped hafnia at orthorhombic polar, monoclinic, and cubic phases in $3\times3\times3$ supercell.}
    \label{fig:polarization}
\end{figure}

In doped hafnia, the oxygen vacancy sites were first restored by placing oxygen atoms at the same positions as in pristine hafnia. All the atomic positions in doped hafnia were adjusted under periodic boundary conditions to closely match the pure structure. Then, we applied same amounts of polarization lattice and Born effective charge polarization quantum identically to the pristine case. Compared with pure hafnia, the electric polarization of the orthorhombic polar phase in doped hafnia decreases with increasing Y concentration, but rises again at higher Y content (Fig. \ref{fig:polarization}). The minimum polarization of 0.43 C/m$^2$ occurs at 11.1\% Y. For the monoclinic and cubic phases, Y doping induces only a relatively small -less than 0.09 C/m$^2$ - polarization. According to the formation energy (Fig. \ref{fig:figure1}(c)), the polar and cubic phase becomes more stable than monoclinic phase as Y dosage reaches 14.8\%. At those high Y concentrations (x =14.8\%, 18.5\%, or 22.2\%), the energetically favored polar and cubic phases can reach \(0.46\sim0.54\) C/m$^2$ and \(0\sim0.06\) C/m$^2$, respectively. So, for hafnia with critical Y doping, considerable electric polarization should be practical in pure polar phase, but mixing with cubic phase will dramatically decrease the electric polarization, which probably explains the small polarization of 0.06 C/m$^2$ measured in previous experiments \cite{xuKineticallyStabilizedFerroelectricity2021}.

Besides, we clarify that in 14.8\% doped hafnia (see Fig. \ref{fig:figure1}(c)), the lower formation energy at orthorhombic polar phase than that at cubic phase doesn't clearly reveal that the polar phase is more stable than the cubic phase, given their small formation energy difference. Those energies are obtained from fully relaxed structures, which are initialized using the SQS configurations via Monte Carlo sampling until pair and triplet probabilities out to 4~\AA\ and 3.8~\AA. We tested larger SQS cutoffs using 14.8\% Y-doped hafnia as an example, including pair and triplet probabilities out to 4~\AA\ and 3.8~\AA, 4.5~\AA\ and 4.3~\AA, 5~\AA\ and 4.8~\AA, respectively (Fig. S4). It turns out the formation energy difference between polar and cubic phases can be either positive or negative, depending on the SQS cutoffs. Nevertheless, the lower formation energies of both polar and cubic phases compared to the monoclinic phase are consistently maintained in these tests. The testing formation energies are marked in Fig. \ref{fig:figure1}(c) as error bars. These results demonstrate the strong competition between polar and cubic phases in doped hafnia.

Next we discuss why we found the stability of ferroelectric hafnia, whereas earlier computations did not \cite{xuKineticallyStabilizedFerroelectricity2021, pavoniEffectDopingMonoclinic2022, batraDopantsPromotingFerroelectricity2017}. First, the previous calculation used 16 formula unit cell with 48 atoms \cite{xuKineticallyStabilizedFerroelectricity2021}, which is too small to simulate a system with the doping concentration as high as 12.5\%. Second, Y-doping is expected to generate oxygen vacancies in hafnia, and the distribution of these vacancies is associated with the substituted Y. Experimental studies have shown that in hafnia thin films, the oxygen migration is even reversible at the interface between the hafnia-based capacitor and the top electrode, regardless of whether the electrode is oxygen reactive or not \cite{nukalaReversibleOxygenMigration2021}. In experiments on hafnia bulk \cite{xuKineticallyStabilizedFerroelectricity2021}, fast quenching is typically used to prevent the diffusion of oxygen vacancies. However, in earlier computational studies \cite{pavoniEffectDopingMonoclinic2022, batraDopantsPromotingFerroelectricity2017}, the oxygen vacancies were either neglected or assumed to be located near the dopants. In contrast, our SQS-based structures allow for greater configurational flexibility. Upon fully structural relaxation, we observe that oxygen vacancies are positioned either close to Y (at distances of approximately 2-2.3~\AA\@) or farther away (at distances greater than 3.5~\AA\@). Crucially, the ferroelectric phase in hafnia exists at high temperatures, which redistributes the oxygen vacancy and promotes a more random spatial distribution. Besides, in previous experiments \cite{musfeldtStructuralPhasePurification2024}, pressure in Y-doped hafnia triggers the phase transition from orthorhombic polar phase to tetragonal phase at 12\% Y-doped hafnia. But the formed tetragonal phase is not metastable upon pressure release, which might be attributed to the high Y concentration. In our calculations (Fig. S2(c)), the tetragonal strain in cubic phase emerges only at low Y concentrations, and disappears when the Y concentration exceeds 11.1\%.

In summary, we find that the orthorhombic polar ferroelectric phase is stabilized by Y$_2$O$_3$-doping in bulk hafnia and zirconia. Further experiments to grow ferroelectric crystals and ceramics are encouraged, possibly leading to a new class of bulk ferroelectrics to join perovskites, with potentially broad uses, especially in extreme environments and at high temperatures, given the thermal and environmental stability of hafnia and zirconia. Besides, we find that the electric polarization of the orthorhombic polar phase in doped hafnia decreases with increasing Y concentration, but rises again as Y concentration reaches 14.8\%. The negative longitudinal piezoelectric effect is maintained in doped hafnia bulk.

\begin{acknowledgments}
This work is supported by U.S. Office of Naval Research Grant No. N00014-20-1-2699 and the Carnegie Institution for Science. Computations were supported by high-performance computer time and resources from the DoD High Performance Computing Modernization Program and Carnegie computational resources. R.E.C. gratefully acknowledges the Gauss Centre for Supercomputing e.V. [https://www.gauss-centre.eu] for funding this project by providing computing time on the GCS Supercomputer SuperMUC-NG at Leibniz Supercomputing Centre (LRZ) [https://www.lrz.de].
\end{acknowledgments}

\section*{DATA AVAILABILITY}
The data that support the findings of this study are available from the corresponding author upon reasonable request. 

\section*{REFERENCES}
\bibliographystyle{aipnum4-1}
\bibliography{zotero}

\end{document}


\title{\textbf{Supplemental Material \\ Stabilization of Ferroelectric Hafnia and Zirconia through Y$_2$O$_3$ doping}}%

\author{Li Yin}
 \email{Author to whom all correspondence should be addressed.\\lyin@carnegiescience.edu}
\author{Cong Liu}
\author{R. E. Cohen}
\affiliation{Extreme Materials Initiative, Earth and Planets Laboratory, Carnegie Institution for Science \\ 5241 Broad Branch Road NW, Washington, District of Columbia 20015, USA}

\maketitle

\section*{\label{sec:level4}Methods}
 Density functional computations were performed with the Vienna Ab-initio Simulation Package (VASP) \cite{kresseEfficiencyAbinitioTotal1996, kresseEfficientIterativeSchemes1996} using revised Perdew–Burke–Ernzerhof for solids (PBEsol) functional in the generalized gradient approximation (GGA) family \cite{perdewGeneralizedGradientApproximation1996, perdewRestoringDensityGradientExpansion2008}. In the PAWs,the valence electron configurations of 5\textit{s}$^2$5\textit{p}$^6$5\textit{d}$^4$, 4\textit{s}$^2$4\textit{p}$^6$4\textit{d}$^3$, and 2\textit{s}$^2$2\textit{p}$^4$ are used in Hf, Y, and O atoms, respectively. The outermost cutoff radius of Hf, Y, and O atoms is 1.27, 1.11, and 0.80 \r{A}, respectively. The plane-wave energy cutoff is set to 475 eV. Structures are fully relaxed with $\Gamma$-only \textbf{k}-mesh and the convergence criteria of $10^{-6}$ eV for energy and 0.001 eV/{\r{A}} for atomic forces. 
 
 The elastic constants and piezoelectric tensors were calculated via the finite differences method in VASP, using an energy cutoff of 600 eV and $4\times4\times4$ \textbf{k}-mesh of the Brillouin zone for pure hafnia (12 atoms). For doped hafnia in $2\times2\times2$ sueprcell (~96 atoms), the $2\times2\times2$ \textbf{k}-grid is used for piezoelectric computations. 
 
 For pure and 6.25\% Y-doped hafnia, the $4\times4\times4$ and $2\times2\times2$ \textbf{k}-meshes of the Brillouin zone are respectively used to calculate their electric polarization through Berry phase method. We have tested $4\times4\times8$ and $2\times2\times4$ \textbf{k}-meshes for calculating electric polarization, which produce the electric polarizations of 0.55 C/m$^2$ and 0.75 C/m$^2$, same with the values under $4\times4\times4$ and $2\times2\times2$ \textbf{k}-meshes. 
 
 We used the special quasiharmonic structure (SQS) method as implemented in the Alloy Theoretic Automated Toolkit \cite{zungerSpecialQuasirandomStructures1990, vandewalleAlloyTheoreticAutomated2002, vandewalleEfficientStochasticGeneration2013} to generate the structures of Y-doped hafnia with oxygen vacancies. Pair clusters with atomic distances up to 4 \r{A} and triplet clusters with atomic distances up to 3.8 \r{A} are considered for searching the best SQS structure. 
 In ferroelectric hafnia bulk (space group \textit{Pca}2$_1$), the atomic distances are 2.0-2.3~\AA\ between Hf and O atoms and 3.3-3.8~\AA\ between Hf atoms. We sampled site positions using Monte Carlo until pair and triplet probabilities out to 4~\AA\ and 3.8~\AA\ were as close as possible to a random solid solution for each composition in each supercell. 

Interatomic force constants for determining vibrational properties were calculated by the supercell approach implemented in Phonopy \cite{togoFirstPrinciplesPhonon2015}, with the energy cutoff of 600 eV and energy convergence criteria of $10^{-6}$ eV. For pure ($2\times2\times2$ supercell) and Y-doped ($1\times1\times1$ supercell) hafnia, the $\Gamma$-centered \textbf{k}-mesh $2\times2\times2$ is utilized for phonon dispersions with the non-analytical term correction \cite{giannozziInitioCalculationPhonon1991, gonzeInteratomicForceConstants1994}. 
 
\section*{Phonons in Y$_2$O$_3$ doped zirconia.}
We computed the phonon dispersion for Y$_2$O$_3$-doped zirconia in the cubic phase. Different from the above disordered SQS structures, we substituted Zr with Y in high-symmetry sites for two small and ordered compositions: Y$_2$Zr$_2$O$_7$ and Y$_2$Zr$_6$O$_{15}$ (corresponding to 50\% and 25\% Y concentrations, respectively). Oxygen vacancies were introduced to maintain charge balance, resulting in the space groups of the two supercells being \textit{Cmmm} and \textit{Amm}2, respectively. The oxygen vacancies are located at (1/4, 1/4, 1/4) and (1/8, 1/4, 1/4) sites, respectively. These Y-doped zirconia are unstable with imaginary frequencies under quasi-harmonic approximation (Fig. S3). Then, we used AIMD for 12 ps and on-the-fly machine learning AIMD for 240 ps to generate two trajectories of Y$_2$O$_3$-doped zirconia solid solution at 300 K with a time step of 1 fs in the NVT ensemble with the Nose-Hoover thermostat. The anharmonic phonon spectrums at 300 K for the two ordered supercells were calculated by the mode decomposition technique using vibrational density of states derived from molecular dynamics (MD) simulations with Dynaphopy code \cite{carrerasDynaPhoPyCodeExtracting2017}. We find these Y-doped zirconia becomes stable -without imaginary phonons- at 300 K (Fig. S3), showing that Y-doping in zirconia dynamically stabilizes the cubic phase at room temperature.

\begin{table}[b]
\caption{\label{tab:tableS1}Computed elastic tensors (\textbf{C}, GPa) and lattice constants \textit{a}, \textit{b}, \textit{c} (\r{A}) in the ferroelectric phase of 0\% (pure), 6.25\%, and 12.5\% Y-doped hafnia at 0 K. The doping structure is in $2\times2\times2$ supercell. Indices are given in Voigt notation.
}
\begin{ruledtabular}
\begin{tabular}{lcccc}
\textrm{Index}&
\textrm{0\% Y in Ref \cite{duttaPiezoelectricityHafnia2021}}&
\textrm{0\% Y}&
\textrm{6.25\% Y}&
\textrm{12.5\% Y}\\
\colrule
C11	&413.6	&407.66	&377.59	&360.54\\
C22	&407.8	&412.57	&395.00	&405.22\\
C33	&394.6	&399.33	&374.94	&382.68\\
C12	&162.3	&161.32	&156.59	&138.33\\
C13	&123.4	&123.12	&124.73	&109.54\\
C23	&132.8	&136.08	&133.21	&131.47\\
C44	&94.4	&94.39	&84.27	&73.09\\
C55	&98	    &96.15	&78.34	&66.66\\
C66	&140.4	&138.93	&124.00	&91.43\\
\\
\textit{a} &5.21 &5.21 &5.22 &5.19\\
\textit{b} &5.00 &5.00 &5.04 &5.07\\
\textit{c} &5.03 &5.03 &5.05 &5.09\\
\end{tabular}
\end{ruledtabular}
\end{table}


\begin{table}[t]
\caption{\label{tab:tableS2}Born effective charge tensors in pure hafnia at ferroelectric \textit{Pca}2$_1$ phase.
}
\begin{ruledtabular}
\begin{tabular}{c c}
\textrm{Z$^*$ for Hf}&
\textrm{Z$^*$ for O}\\
\colrule
$\begin{bmatrix}
5.20419 & -0.00213 & -0.01021 \\
-0.25930 & 5.53620 & -0.14003 \\
0.04243 & -0.20458 & 5.06569
\end{bmatrix}$
&
$\begin{bmatrix}
-2.50269 & -0.93289 & 0.33844 \\
-0.85692 & -3.02951 & -0.71037 \\
0.30847 & -0.66555 & -2.54578
\end{bmatrix}$ \\
\end{tabular}
\end{ruledtabular}
\end{table}


\begin{figure}
    \centering
    \includegraphics[width=0.8\linewidth]{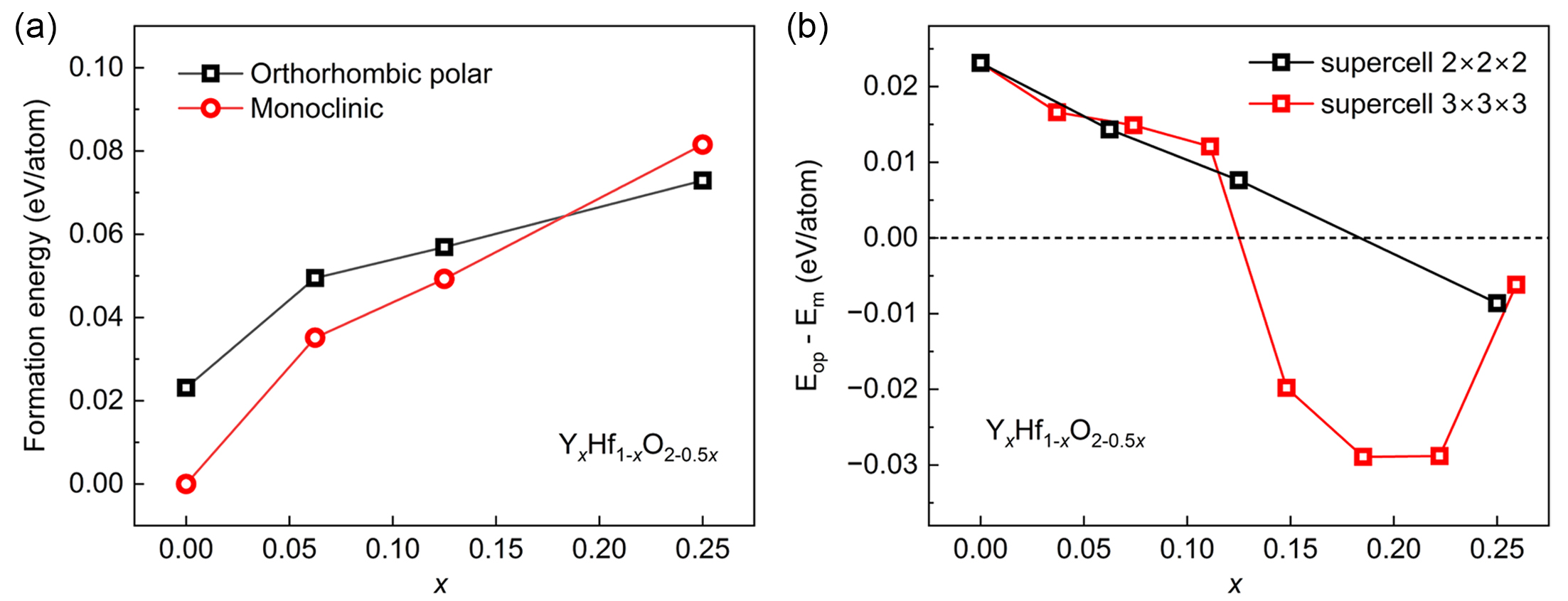}
    \caption{(a) Formation energies, (b) energy differences between phases in a series of fully relaxed Y-doped hafnia in $2\times2\times2$ supercell. The \textit{op} denotes orthorhombic polar phase, and \textit{m} indicates the monoclinic phase.}
    \label{fig:S1}
\end{figure}

\begin{figure}
    \centering
    \includegraphics[width=0.95\linewidth]{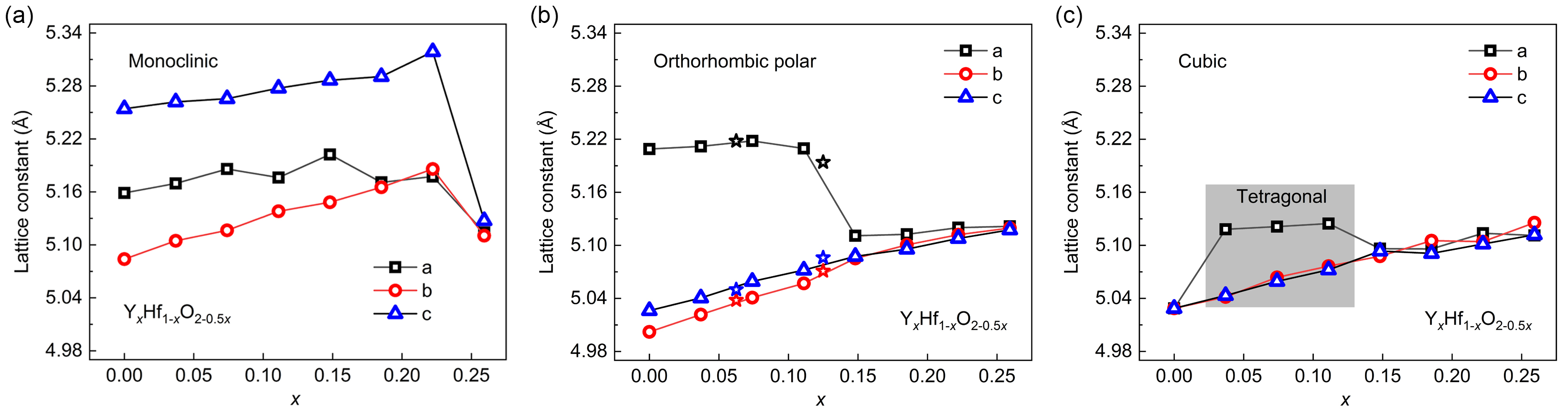}
    \caption{Lattice constants in a series of $3\times3\times3$ Y-doped hafnia supercell fully relaxed from (a) monoclinic, (b) orthorhombic polarized, and (c) cubic phases, respectively. The star symbols in (b) denote the lattice constants of 6.25\% and 12.5\% Y-doped hafnia in $2\times2\times2$ supercell, corresponding to Y-doped structure displayed in Table I and Fig. 3.}
    \label{fig:S2}
\end{figure}

\begin{figure}
    \centering
    \includegraphics[width=0.8\linewidth]{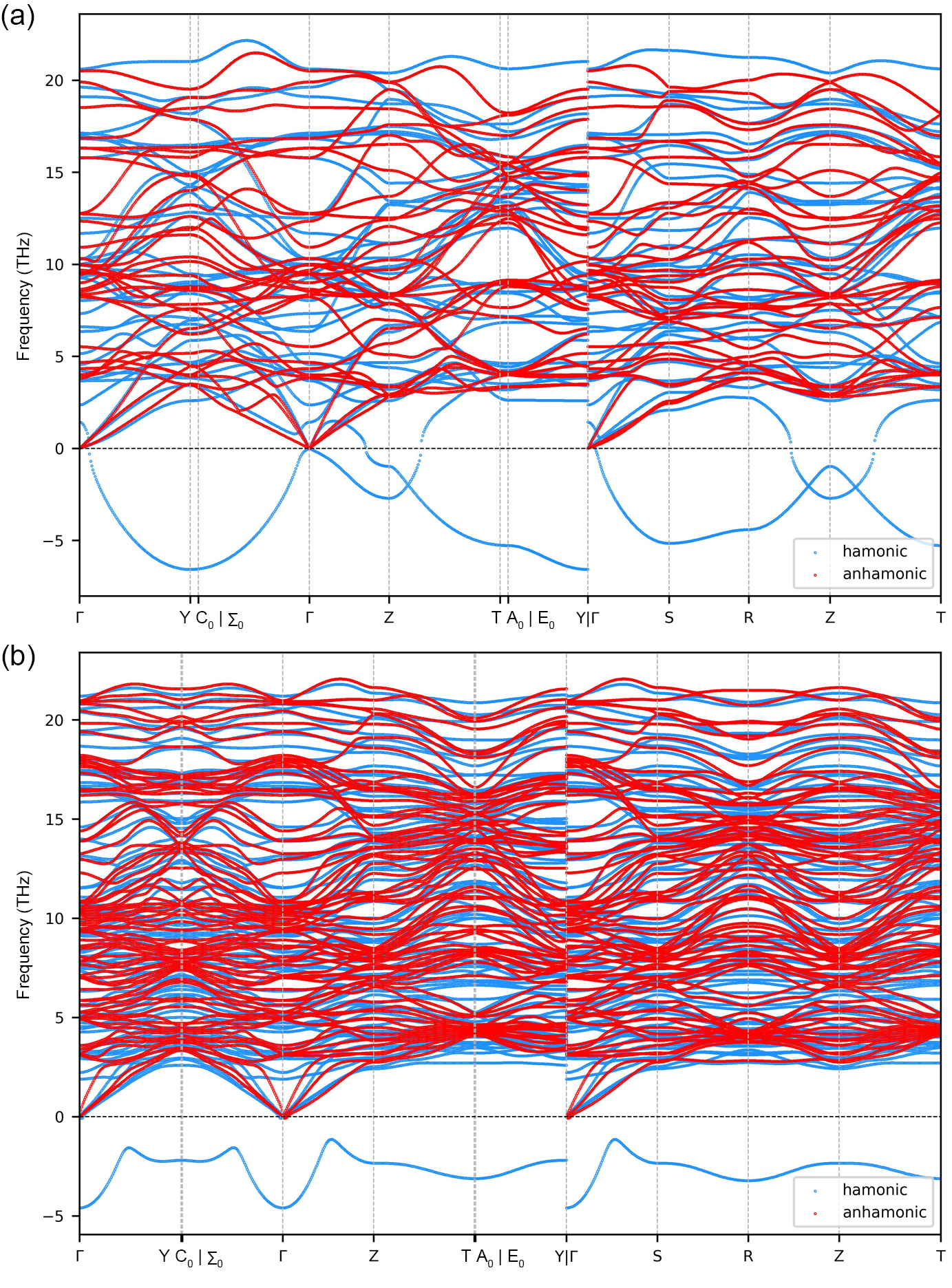}
    \caption{Phonon spectrum with harmonic (0 K) and anharmonic (300 K) effects in (a) 50\% and (b) 25\% Y-doped zirconia at cubic phase, with compositions of Y$_2$Zr$_2$O$_7$ and Y$_2$Zr$_6$O$_{15}$ respectively.}
    \label{fig:S3}
\end{figure}

\begin{figure}
    \centering
    \includegraphics[width=0.8\linewidth]{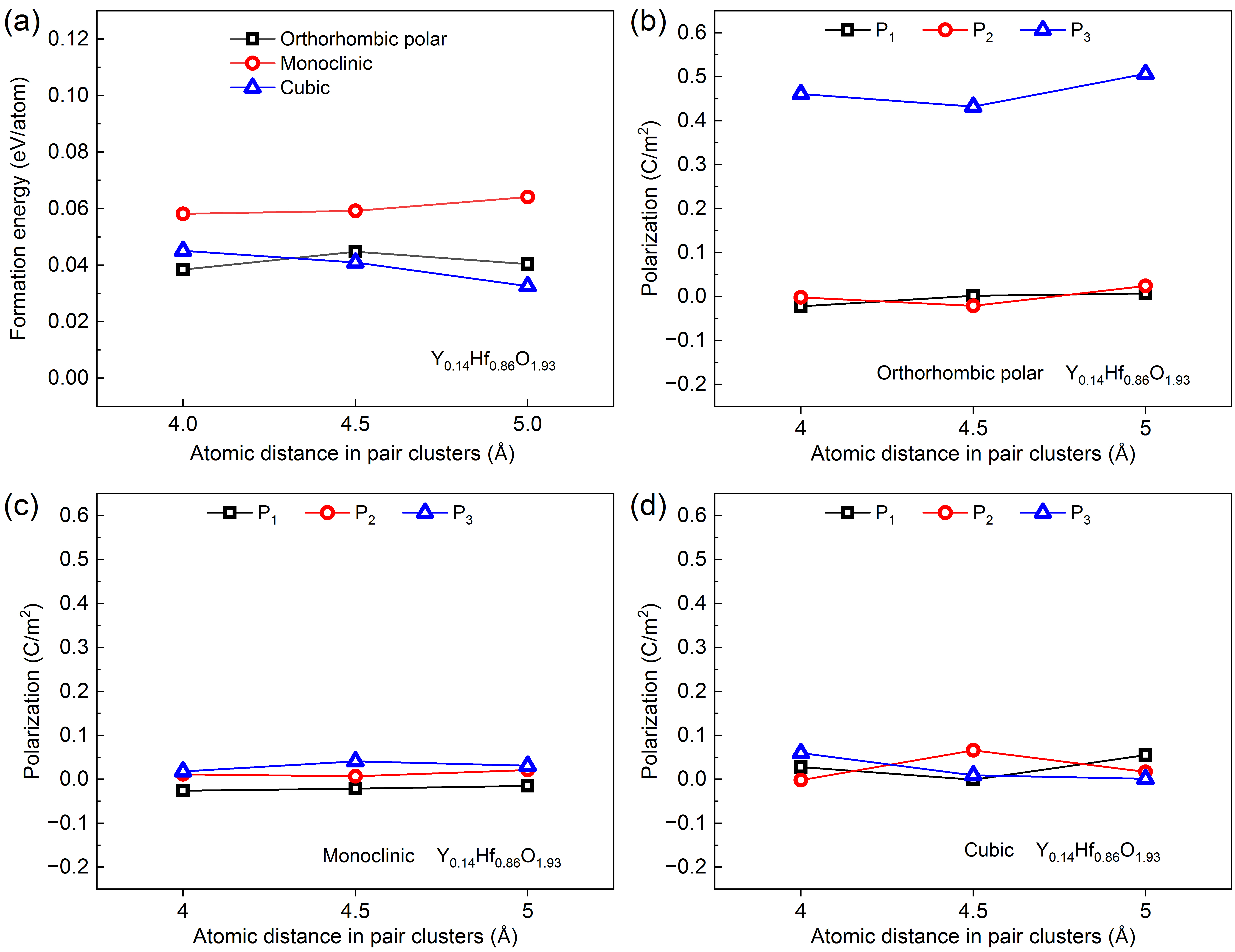}
    \caption{(a) Formation energies and (b) electric polarization in doped hafnia Y$_{0.14}$Hf$_{0.86}$O$_{1.93}$ based on different SQS cutoffs, including pair and triplet probabilities out to 4~\AA\ and 3.8~\AA, 4.5~\AA\ and 4.3~\AA, 5~\AA\ and 4.8~\AA, respectively.}
    \label{fig:SQStest}
\end{figure}

\clearpage
\bibliographystyle{aipnum4-1}
\bibliography{zotero}